\documentclass[prb,preprint,showkeys,floatfix,nofootinbib,superscriptaddress]{revtex4} 

\usepackage{amsmath}  % needed for \tfrac, \bmatrix, etc.
\usepackage{amsfonts} % needed for bold Greek, Fraktur, and blackboard bold
\usepackage{graphicx} % needed for figures

\usepackage{amsthm}
\usepackage{wasysym}
\usepackage{array}

\usepackage{url}

\def\units#1{\hbox{$\,{\rm #1}$}}

\begin{document}

\title{Measurement of the ratio h/e with a photomultiplier tube and a set of LEDs}

\author{F.~Loparco}
\email{francesco.loparco@ba.infn.it} 
\affiliation{Dipartimento di Interateneo Fisica ``M. Merlin'' 
dell'Universit\`a degli Studi e del Politecnico di Bari,
I-70126, Bari, Italy} 
\affiliation{Istituto Nazionale di Fisica Nucleare Sezione di Bari, I-70126, Bari, Italy}

\author{M.~S.~Malagoli}
\affiliation{Dipartimento di Interateneo Fisica ``M. Merlin'' 
dell'Universit\`a degli Studi e del Politecnico di Bari,
I-70126, Bari, Italy} 

\author{S.~Rain\`o}
\affiliation{Dipartimento di Interateneo Fisica ``M. Merlin'' 
dell'Universit\`a degli Studi e del Politecnico di Bari,
I-70126, Bari, Italy} 
\affiliation{Istituto Nazionale di Fisica Nucleare Sezione di Bari, I-70126, Bari, Italy}

\author{P.~Spinelli}
\affiliation{Dipartimento di Interateneo Fisica ``M. Merlin'' 
dell'Universit\`a degli Studi e del Politecnico di Bari,
I-70126, Bari, Italy} 
\affiliation{Istituto Nazionale di Fisica Nucleare Sezione di Bari, I-70126, Bari, Italy}


\date{\today}

\begin{abstract}

We propose a laboratory experience aimed at undergraduate physics students to 
understand the main features of the photoelectric effect and to perform a measurement 
of the ratio $h/e$, where $h$ is the Planck's constant and $e$ is the electron charge. 
The experience is based on the method developed by Millikan 
for his measurements on the photoelectric effect in the years from 1912 to 1915.
The experimental setup consists of a photomultiplier tube (PMT) equipped with a voltage divider 
properly modified to set variable retarding potentials between the photocathode and the first dynode, 
and a set of LEDs emitting at different wavelengths. The photocathode is illuminated with the various 
LEDs and, for each wavelength of the incident light, the output anode current is measured as a function 
of the retarding potential applied between the cathode and the first dynode. 
From each measurement, a value of the stopping potential for the anode current is derived. 
Finally, the stopping potentials are plotted as a function 
of the frequency of the incident light, and a linear fit is performed. 
The slope and the intercept of the line allow respectively to evaluate the ratio 
$h/e$ and the ratio $W/e$, where $W$ is the work function of the photocathode.

\end{abstract}

\maketitle % title page is now complete

\section{Introduction} 
\label{sec:intro}

The Planck's constant $h$ plays a central role in quantum mechanics. It was first introduced 
in 1900 by Max Planck in his study on the blackbody radiation~\cite{planck} as the 
proportionality constant between the minimal increment of energy of a charged oscillator
in a cavity hosting blackbody radiation and the frequency of its associated electromagnetic
wave. In 1905 Albert Einstein explained the photoelectric effect postulating that luminous 
energy can be absorbed or emitted only in discrete amounts, called quanta~\cite{einstein}. 
The light quantum behaved as an electrically neutral particle and was called ``photon''.
The Planck-Einstein relation, $E=h\nu$, connects the photon energy with its associated 
wave frequency.

Nowadays, the measurement of the Planck's constant is ordinarily performed by physics 
students in many educational laboratories, both in universities and in high schools. 
The most common techniques exploit the blackbody radiation 
(see refs.~\cite{george,manikopoulos,crandall,dryzek,brizuela}), the
emission of light by LEDs when a forward bias is applied
(see ref.~\cite{nieves}) or the photoelectric effect
(see refs.~\cite{oleary,hall,bobst,barnett,garver}).

Almost all measurements of $h$ exploiting the photoelectric effect are based on the principle
of the experiment carried out by Millikan in the years from 1912 to 1915~\cite{millikan}. 
It is worth here to point out that, although the title of his 1916 article is 
``A direct photoelectric determination of Planck's $h$'', in his experiment Millikan could 
not measure $h$, but he measured the ratio $h/e$ between 
the Planck's constant and the electron charge; then, using the value of $e$ that he had previously
measured~\cite{millikan2,millikan3}, he was able to evaluate $h$
\footnote{When discussing his
results with sodium, Millikan writes:
\begin{quote}
``We may conclude then that the slope of the volt-frequency line for sodium is
the mean of $4.124$ and $4.131$, namely $4.128 \times 10^{-15}$ which, with my
value of $e$, yields $h=6.569 \times 10^{-27} \units{erg~sec}$''. 
\end{quote}
}.

Although the apparatus used by Millikan was rather complex, the method chosen for 
the measurement of $h/e$ is quite simple.
The detectors basically consisted of a metal surface, which was illuminated with different 
monochromatic light sources, and a collector electrode, kept at a lower potential with respect 
to the metal. For each frequency of the incident light, the potential was adjusted 
until no current flowed through the collector, thus allowing to evaluate the ``stopping 
potential''. It is straightforward to show that the stopping potential increases
linearly with the frequency of the incident light, and the slope of the straight line is
given by $h/e$. A linear fit of the stopping potentials 
at different frequencies allows therefore to evaluate Planck's constant
if the value of the electric unit charge $e$ is known.

In the various didactic experiments proposed to measure the ratio $h/e$ exploiting
photoelectric effect, a variety of devices and light sources are used (see again the examples
in refs.~\cite{oleary,hall,bobst,barnett,garver}). 
In this paper we present a novel didactic experience for measuring $h/e$ using
a photomultiplier tube (PMT) and a set of light emitting diodes (LEDs). We propose this
experience to undergraduate physics students attending our introductory laboratory course
to quantum physics. PMTs are very common devices in atomic and nuclear physics, and can be
easily available in a didactic laboratory. The main advantage of a PMT with respect to a
conventional photoelectric cell resides in the fact that photoelectrons extracted at 
the cathode are considerably amplified (the typical gain is of $\sim 10^{5} \div 10^{6}$), 
thus providing detectable output currents even when a large fraction of them is 
repelled back to the photocathode. This feature will help in evaluating the stopping potential
as we will discuss later in Sec.~\ref{sec:analysis}.

The paper is organized as follows: in Sec.~\ref{sec:setup}
we describe the instrumentation and the theoretical principles of the measurement;
in Sec.~\ref{sec:analysis} we propose a method to analyze the data collected in the 
experiment; finally in Sec.~\ref{sec:discussion} we discuss the results obtained and
some possible strategies to improve the experiment.

\section{Experimental setup}
\label{sec:setup}

PMTs are widely used in many fields of physics to convert an incident flux of light into an electric signal. 
A PMT is a vacuum tube consisting of a photocathode and an electron multiplier, composed by a set of electrodes 
called ``dynodes'' at increasing potentials. Incident light enters into the tube through the photocathode, 
and the electrons are extracted as a consequence of photoelectric effect (photoelectrons). Photoelectrons 
are accelerated by an appropriate electric field towards the first dynode of the multiplier, 
where a few secondary electrons are extracted. The multiplication process is repeated through all 
the dynodes, until the electrons ejected from the last dynode are finally collected by the anode, 
which produces the current signal that can be read out. A PMT can be operated either in pulsed mode 
or with a continuous light flux.

In our experience we used a Philips XP 2008 PMT~\cite{philips}. The photocathode is a thin film 
(a few $\units{nm}$ thick) made of a Sb-Cs alloy deposited over a glass window, and is sensitive 
to a range of wavelengths that extends from approximately $280 \units{nm}$ to $700 \units{nm}$. 
The upper limit of this interval is set by the work function of the metal alloy, while the lower 
limit is set by the glass of the window, which is opaque to UV photons. The photocathode works 
in transmission mode, i.e. photoelectrons are collected from the opposite side of incident light. 
The electron multiplier consists of a set of $10$ dynodes, each made of a Be-Cu alloy. 

\begin{figure}[t]
\includegraphics[width=0.95\columnwidth]{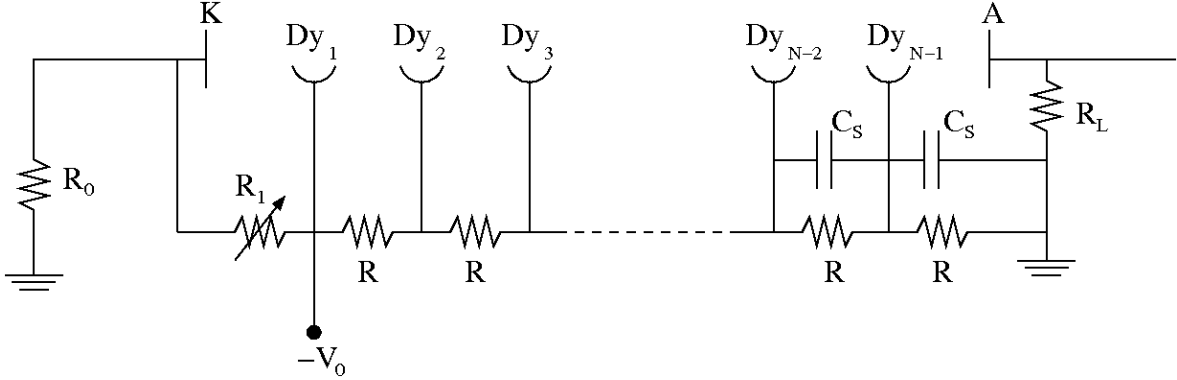}
\caption{Electric scheme of the voltage divider of the Philips XP 2008 PMT used in the experiment.
The photocathode is indicated with K, the anode with A and the various dynodes with \mbox{Dy$_i$}. 
The values of the resistances are $R_{0}=1\units{M\Omega}$, $R=150\units{k\Omega}$ and $R_{L}=100\units{k\Omega}$.
The resistance $R_{1}$ of the potentiometer connecting the cathode with the first dynode is allowed 
to vary in the range $[0,10]\units{k\Omega}$. The capacitors in the last stages are used to keep
the voltage differences stable, and their capacitance is $C_{S}=0.01\units{nF}$.}
\label{fig:divider}
\end{figure}

Fig.~\ref{fig:divider} shows the electric scheme of the voltage divider used to provide the voltage 
differences to the dynodes. Unlike a standard PMT voltage divider, here the negative high voltage is 
supplied to the first dynode (not to the photocathode, which is grounded through the resistor $R_{0}$), 
thus ensuring that the photocathode K is 
kept at higher voltage with respect to the first dynode Dy1. The voltage difference between the two 
electrodes can be adjusted by changing the variable resistance $R_{1}$, and is given by:

\begin{equation}
 V_{K} - V_{Dy1} = \frac{R_{1}V_{0}}{R_{0}+R_{1}} \approx \frac{R_{1}}{R_{0}}V_{0}.
\label{eq:vkd}
\end{equation}
In writing eq.~\ref{eq:vkd} we took into account the fact that $R_{1} \ll R_{0}$ (see Fig.~\ref{fig:divider}). 
If a high voltage $V_{0}=1000\units{V}$ is supplied to the PMT, a maximum voltage difference 
of $10\units{V}$ can be applied between K and Dy1. Since $V_{K} > V_{Dy1}$, photoelectrons extracted 
from K will be slowed down in their motion towards Dy1 
and eventually sent back to K. On the other hand, like in standard PMT voltage dividers, the dynodes 
and the anode are kept at increasing potentials (if $V_{0}=1000\units{V}$, the average voltage 
differences between each pair of dynodes will be of the order of $100\units{V}$). 
In this way, photoelectrons eventually reaching Dy1 will be multiplied, producing a detectable 
current signal at the anode A, and consequently a voltage difference across the load resistor $R_{L}$.

\setlength{\tabcolsep}{12pt}

\begin{table}[t]
 \begin{tabular}{lcc}
  \hline 
  LED & Peak wavelength ($\units{nm}$) & Peak frequency ($\units{10^{14} Hz}$) \\ 
  \hline	
  red    & $631 (17)$ & $4.75 (0.13)$ \\
  yellow & $585 (15)$ & $5.11 (0.13)$ \\
  green  & $563 (12)$ & $5.32 (0.11)$ \\
  blue   & $472 (14)$ & $6.34 (0.18)$ \\
  violet & $403 (6)$  & $7.42 (0.08)$ \\
\hline
 \end{tabular}
\caption{Main features of the emission spectra of the LEDs used in the measurement. The wavelength and frequency
spectra $(dN/d\nu \propto (1/\nu^{2}) dN/d\lambda)$ have been fitted with gaussian functions. Here we
report the mean values and, in brackets, the standard deviations of each fit function.}
\label{tab:LED}
\end{table}

To perform our measurements we used five LEDs, emitting visible light of different colors ranging 
from red to violet. We preliminarily measured their emission spectra using an OCEAN OPTICS HR2000+ 
spectrometer~\cite{ocean}. 
Tab.~\ref{tab:LED} shows the peak values of the wavelengths and frequencies of each LED. 
The emission spectra of each LED have been fitted with gaussian functions. 
The values of the peak wavelengths (frequencies) and the corresponding standard
deviations are reported in Tab.~\ref{tab:LED}.

\begin{figure}[t]
\includegraphics[width=0.95\columnwidth]{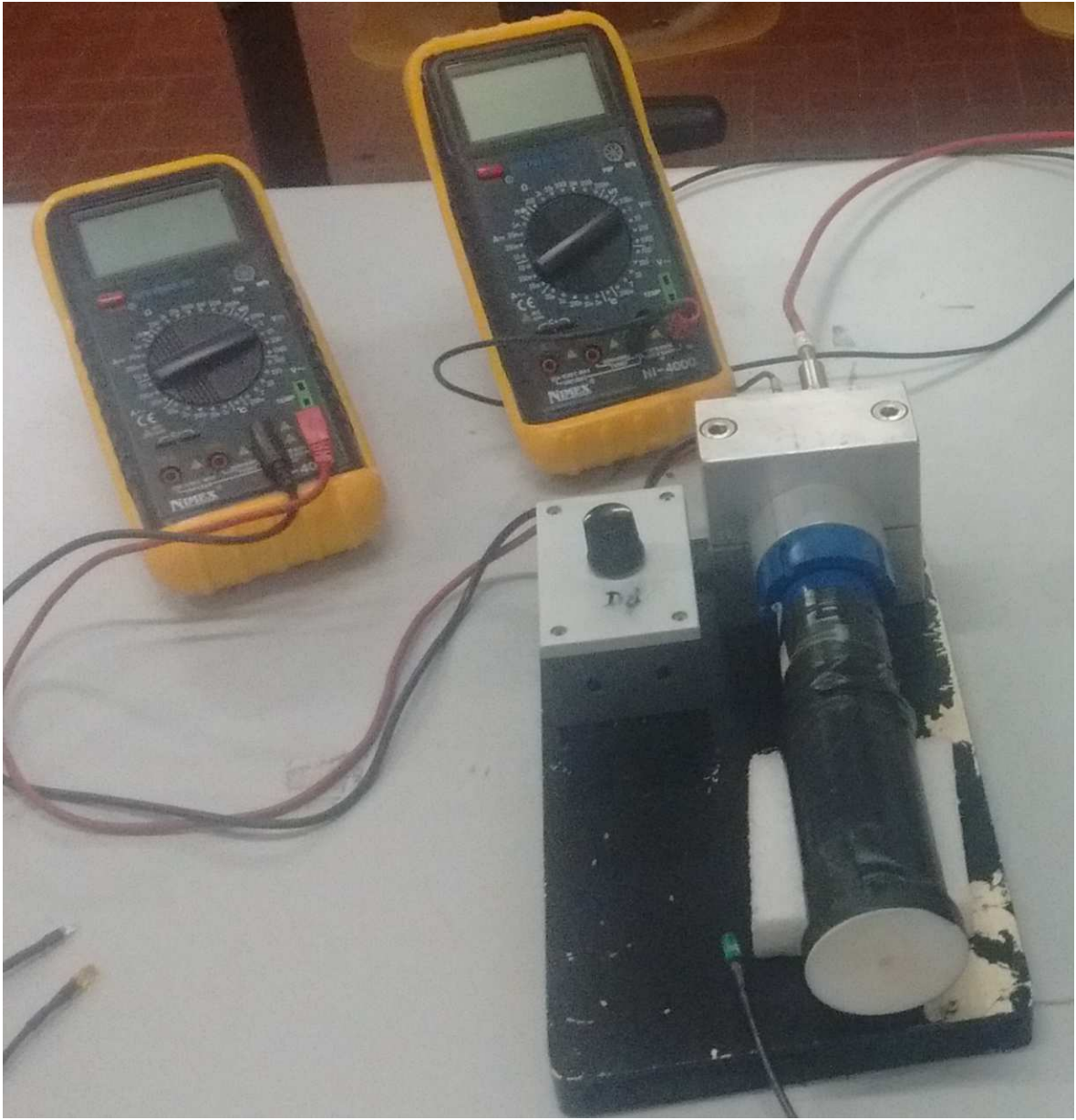}
\caption{Photo of the experimental setup. The window of the photocathode 
is coupled to a plastic support with a hole drilled at its center, in which the different 
LEDs can be inserted. The PMT and the support are wrapped with black tape, to prevent 
external light entering in the device. Two digital multimeters are used to measure the
voltage differences between the photocathode and the first dynode and across the load resistor.
The knob on the left of the PMT is connected to a potentiometer, which allows to
adjust the voltage of the first dynode. 
}
\label{fig:setup}
\end{figure}

Fig.~\ref{fig:setup} shows the experimental setup. The photocathode window is coupled to a plastic 
support with a hole drilled at its center where the different LEDs can be inserted. The PMT and the support 
are wrapped with black tape, to prevent external light entering into the device. The hole is also covered with 
black tape when a LED is inserted to perform a measurement. The voltage differences between K and Dy1 and 
across the load resistor $R_{L}$ are measured by two digital multimeters. The knob placed on the left of the 
PMT is connected to a potentiometer which allows the user to adjust the value of $R_{1}$ and consequently
the voltage $V_{K}-V_{Dy1}$. The high voltage is supplied to the PMT by means of a CAEN N471A 
NIM power supply module~\cite{caen} (not shown in the figure). In our measurements we operated the PMT 
with high voltages in the range $700-1000\units{V}$. This choice allows to keep a high PMT gain without
incurring saturation effects due to large number of electrons flowing across the last dynodes.

The students should investigate the dependence of the voltage difference $V_{L}$ across the load resistor  
on the retarding potential $V_{R}=V_{K}-V_{Dy1}$ for the various LEDs. The value of $V_{L}$  
is proportional to the anode current and consequently to the rate of photoelectrons collected by Dy1. 
During a measurement, the voltage across the LED must be kept constant, thus ensuring that the intensity 
of the light entering the PMT is also constant. 

Photoelectrons extracted from the photocathode will have different initial kinetic energies up to a maximum 
value given by:

\begin{equation}
 E_{K,max} = h \nu - W
\end{equation}
where $h \nu$ is the energy of incident photons and $W$ is the work function of the photocathode. 
If $V_{R}=0$, all the photoelectrons extracted from K will be able to reach Dy1, and a current will flow 
through $R_{L}$. If $V_{R}$ is increased, only the more energetic photoelectrons will be collected by Dy1 
and therefore the output current will decrease. When $eV_{R} \geq E_{K,max}$ the photoelectrons will not 
be allowed to reach Dy1 and the current flowing through $R_{L}$ is expected to vanish. The value 

\begin{equation}
V_{S} = \frac{E_{K,max}}{e} = \frac{h \nu - W}{e} 
\label{eq:vs}
\end{equation}
represents the stopping potential, that depends on the energy of incident photons and on the work function 
of the photocathode. 

From the plots of $V_{L}$ as a function of $V_{S}$ (hereafter we will refer to these plots as 
``photoelectric curves''), the students will be able to evaluate the stopping potential $V_{S}$ for each LED. 
The values of $V_{S}$ will then be plotted against the frequency $\nu$ of the incident light, 
and the data will be fitted with a straight line. According to equation~\ref{eq:vs}, the value of 
$h/e$ will be derived from the slope of the line, while the value of $W/e$ will be derived
from the intercept.\footnote{The intercept corresponds to the ratio
$W/e$ with a change of sign. If voltages are measured in units of $\units{V}$, the value of $W/e$ will 
also be in units of $\units{V}$, and will correspond to the value of $W$ in units of $\units{eV}$.}

\section{Data analysis}
\label{sec:analysis}

\begin{figure}[t]
\includegraphics[width=0.47\columnwidth]{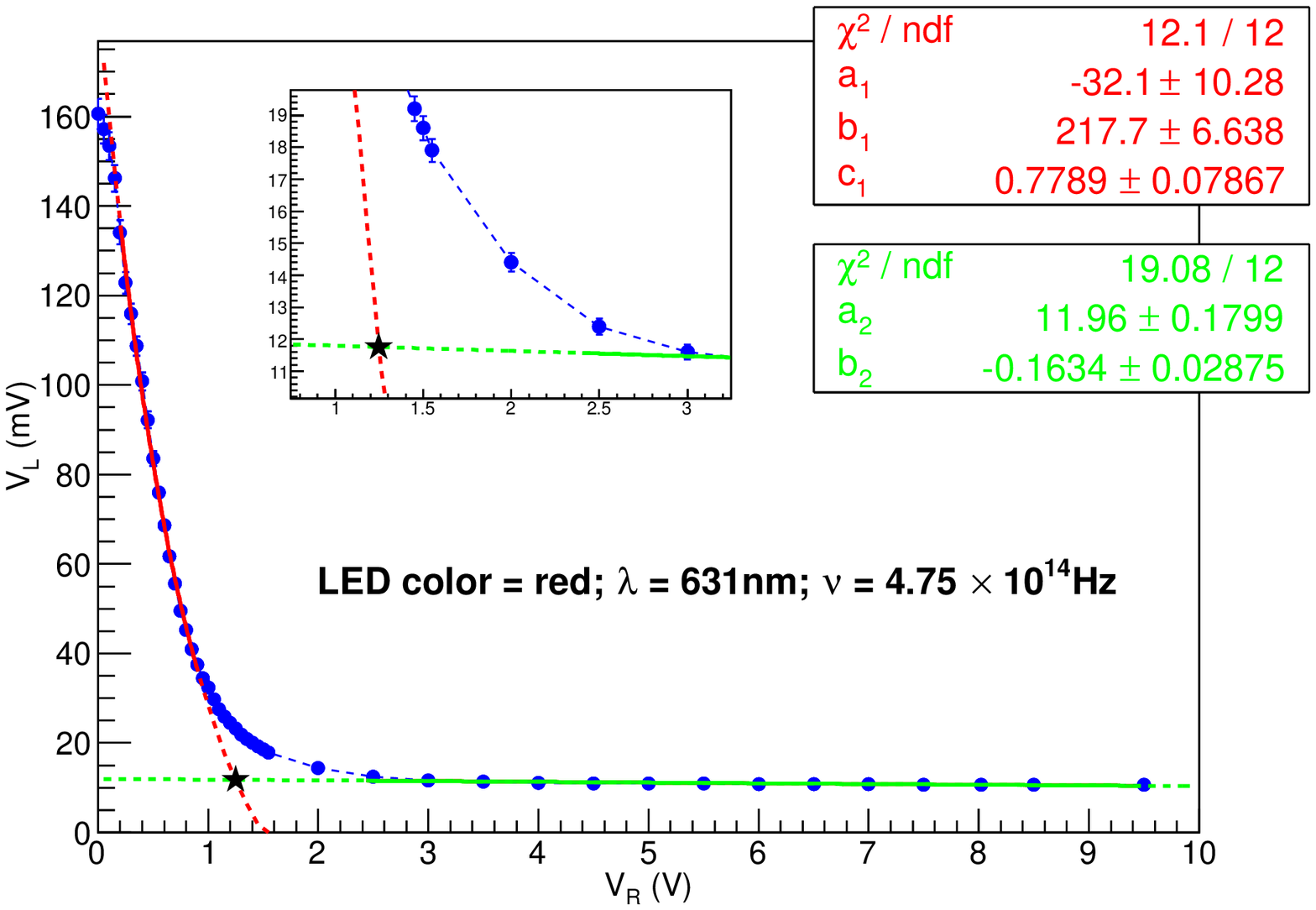}
\includegraphics[width=0.47\columnwidth]{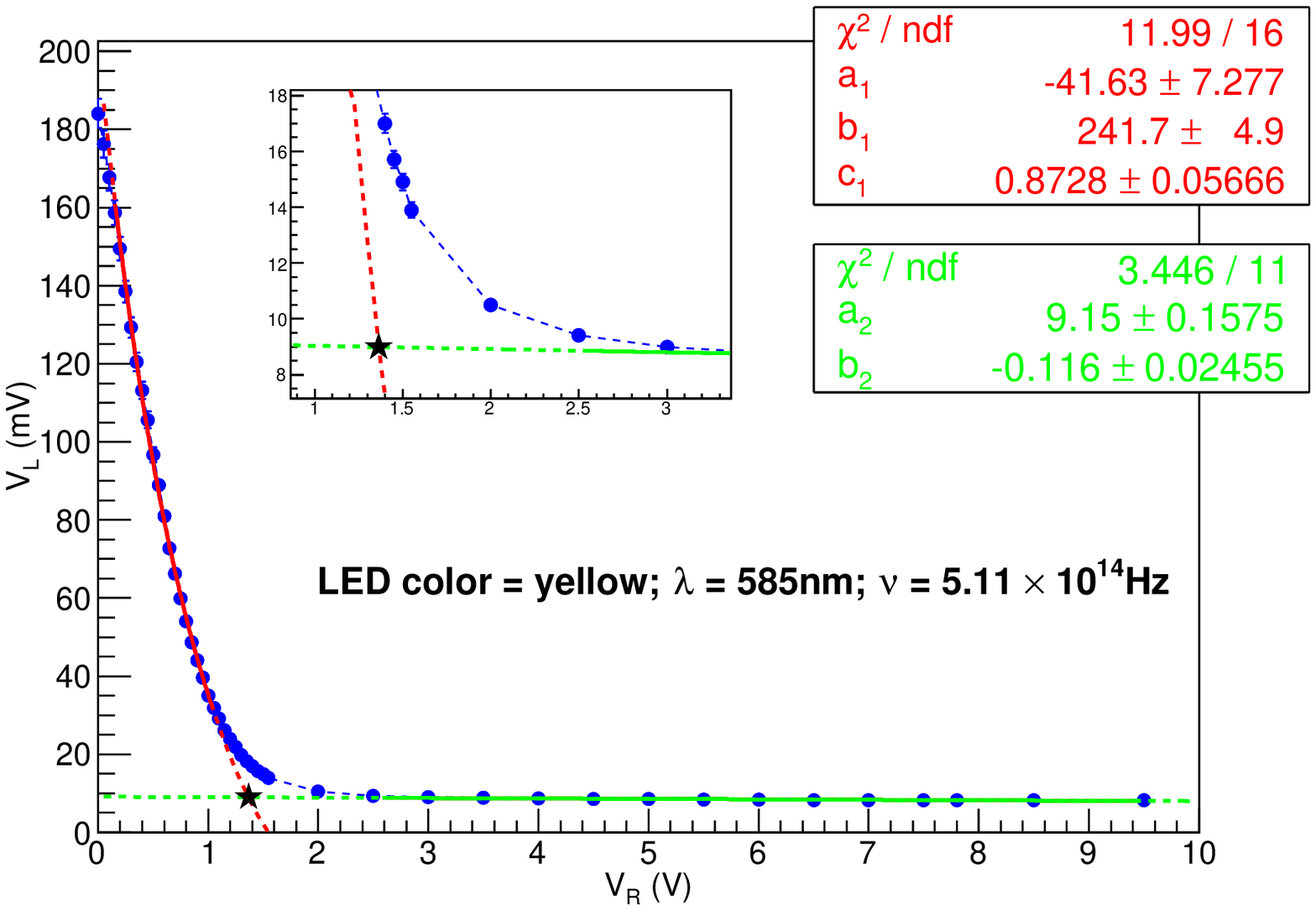}
\includegraphics[width=0.47\columnwidth]{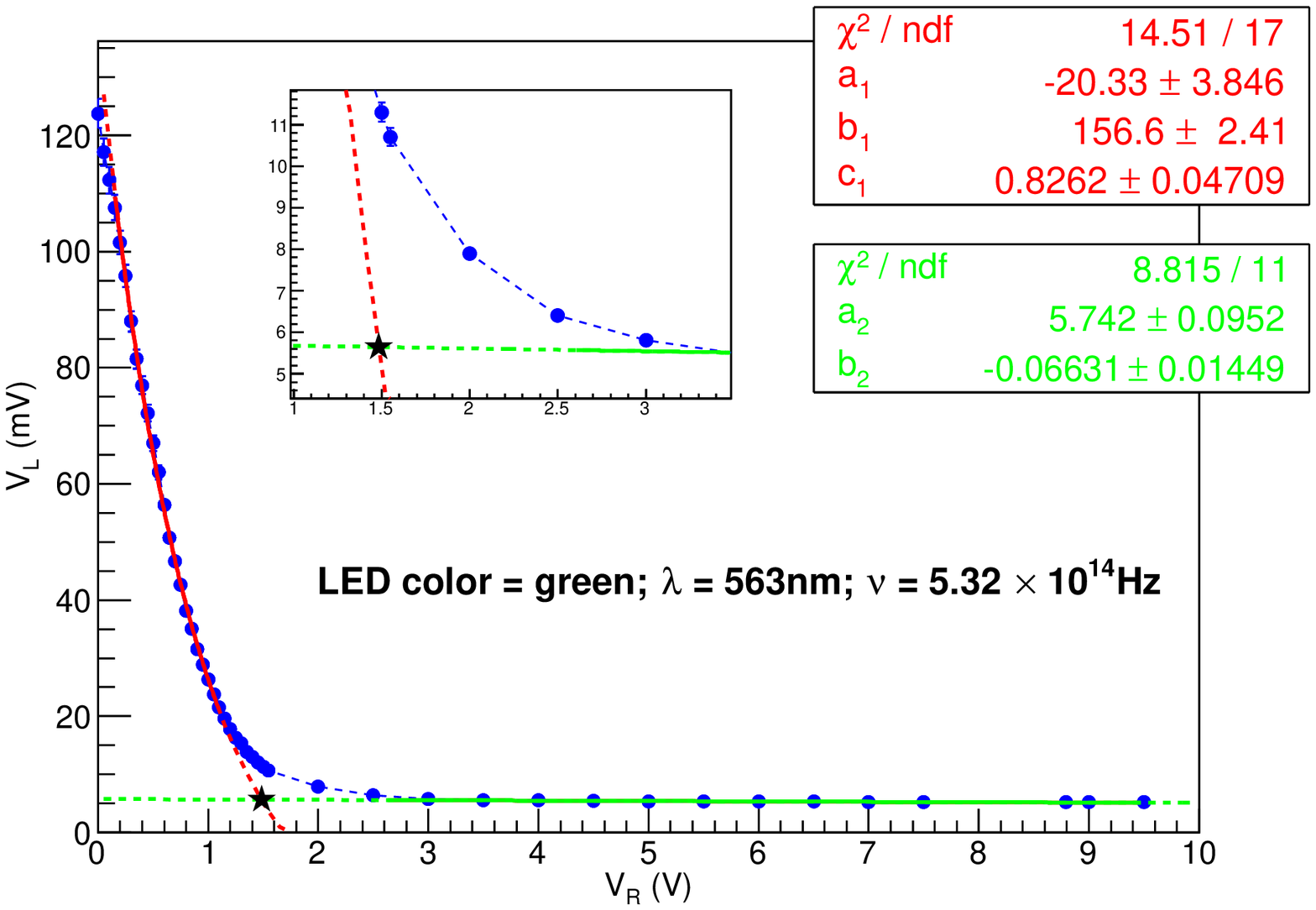}
\includegraphics[width=0.47\columnwidth]{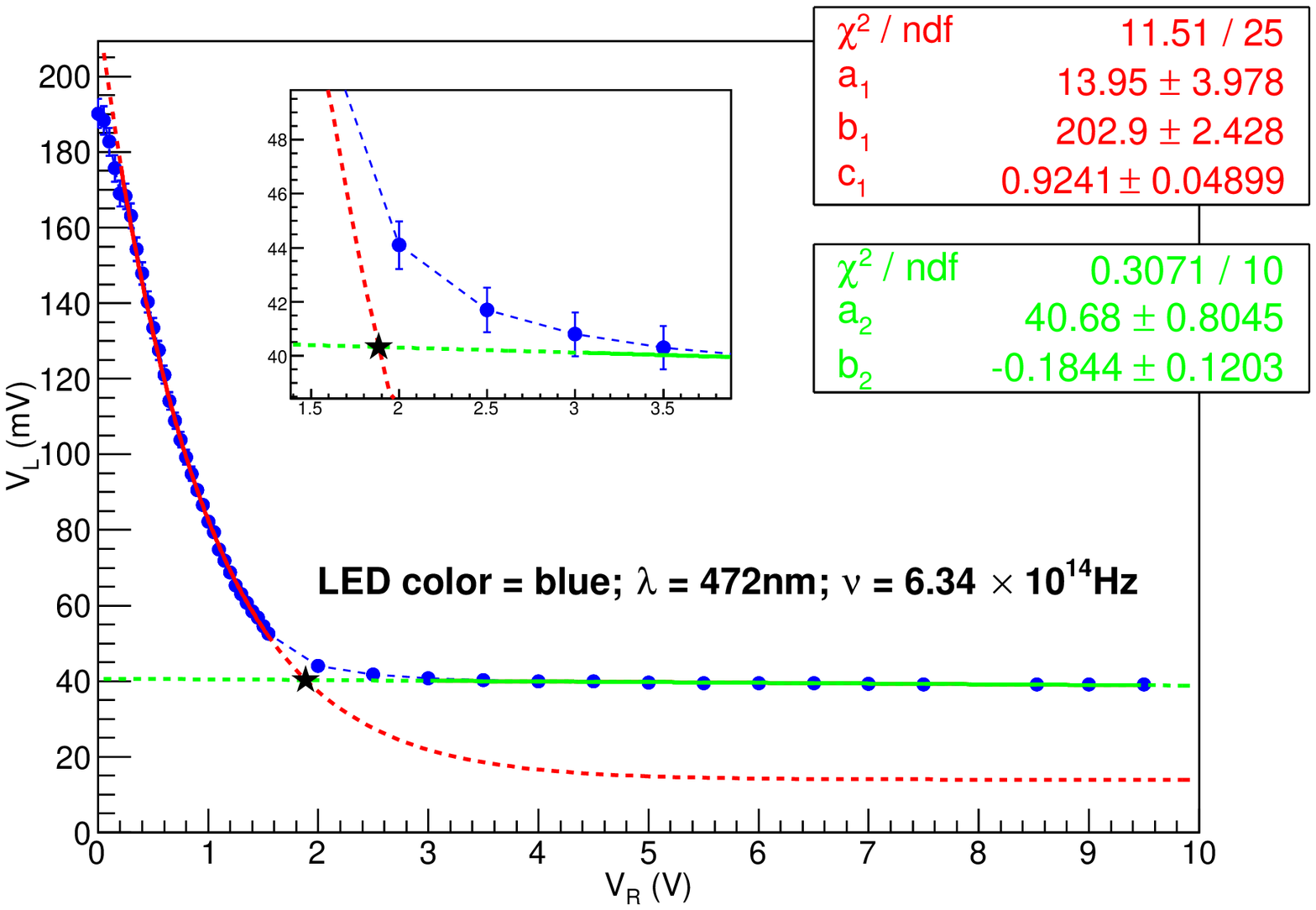}
\includegraphics[width=0.47\columnwidth]{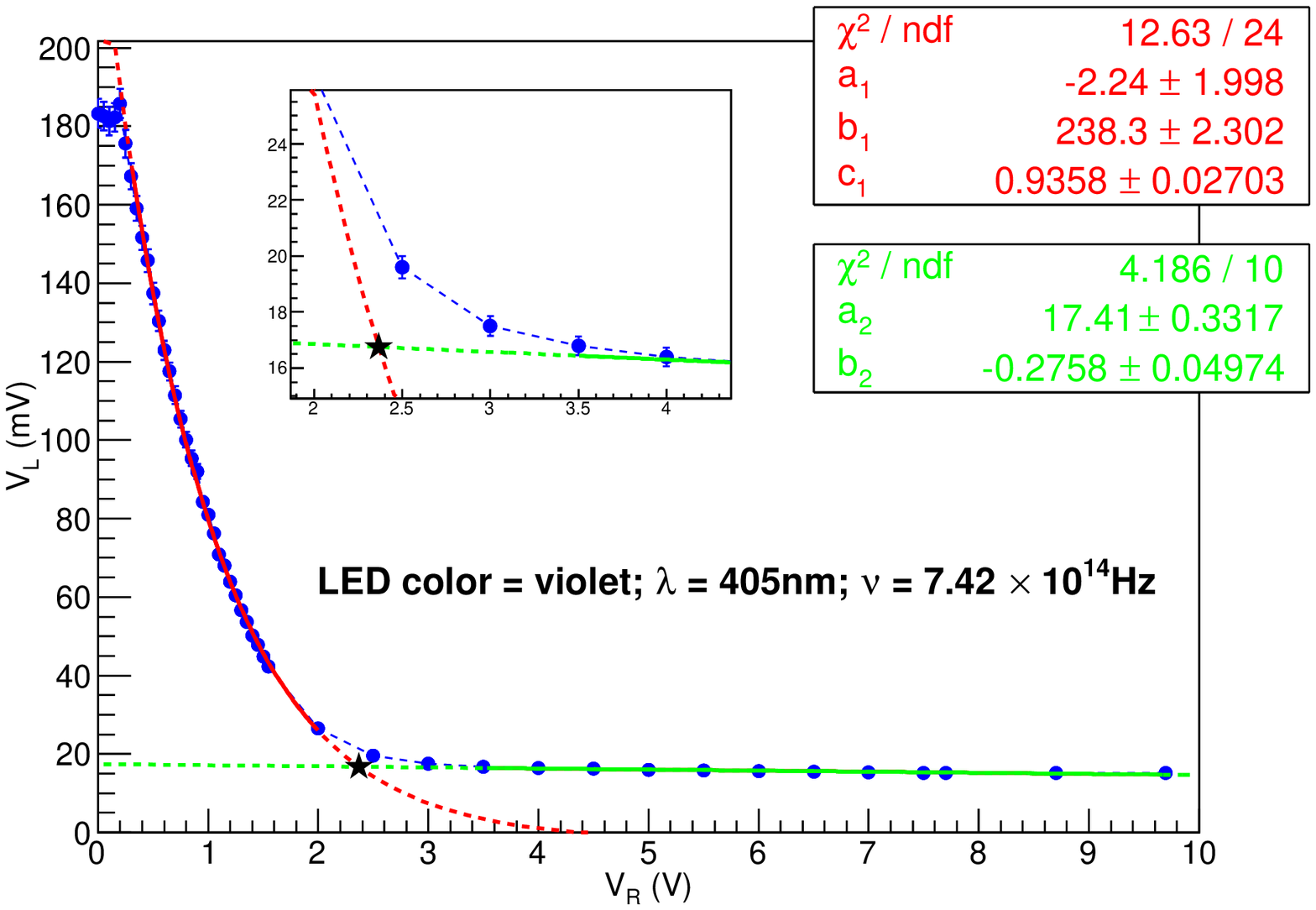}
\caption{Examples of photoelectric curves obtained with the various LEDs. 
The continuous red and green lines represent the fits of the regions 
$V_{R}<V_{S}$ and $V_{R}>V_{S}$ with the two functions in eq.~\ref{eq:fit}. 
The values of the fitted parameters and of the $\chi^{2}$ are shown in the 
top right panels of each plot, with the same color code. 
The dashed lines are obtained extrapolating the two fit functions
outside the corresponding fit regions. A black star is drawn at
the intersection point between the two fit functions. The abscissa of
the intersection point provides the estimate of the stopping potential.
A zoom of the region where the stopping potential is found 
is shown in the inset of each plot.  
}
\label{fig:photocurves}
\end{figure}

Fig.~\ref{fig:photocurves} shows some examples of photoelectric curves obtained when the PMT is illuminated 
with the various LEDs. As expected, the value of $V_{L}$ decreases with increasing $V_{R}$, but 
never drops to zero. This behavior can be explained by taking into account that a fraction 
of the incident photons can pass through the photocathode without interacting, and can extract 
photoelectrons from the first dynode. These electrons are accelerated towards Dy2, 
thus contributing to the output signal because of the high PMT gain. 
Hence, even when $V_{R}>V_{S}$, a background current will 
flow through the load resistor $R_{L}$, and consequently a steady positive value of $V_{L}$ will be measured. 
The fraction of photons extracting photoelectrons from the first dynode changes with the photon energy, 
as the absorption probabilities in the photocathode and in the first dynode are strongly dependent on the photon energy. 
Another possible contribution to the background anode current could be due to ambient light 
entering into the device, but we ruled out this contribution performing a preliminary set of measurements 
with the LEDs being turned off, in which we observed $V_{L}=0$ for any value of $V_{R}$. 
Finally, it is also worth to point out here that the electron optics of a PMT is designed 
for electrodes kept at increasing potentials. Therefore electrons emitted from the photocathode are 
accelerated towards the first dynode and are focused onto its center regardless their emission angle, 
thus ensuring optimal collection efficiency. Setting in our device a retarding potential 
between K and Dy1, we introduce a distortion in the electron optics of the PMT, that affects the 
trajectories of photoelectrons preventing them to reach the first dynode. However, 
even when $V_{R}>V_{S}$, some photoelectrons travelling in weaker field regions might be able 
to reach the first dynode, contributing to the output signal.   

We performed several sets of measurements, changing either the high voltage $V_{0}$ supplied 
to the PMT or the intensity of the light emitted by the various LEDs. An increase of $V_{0}$ will
result in an increase of the gain of the electron multiplier, while an increase of the light 
intensity will result in an increase of the number of photoelectrons. In particular, we observe 
that, if the light intensity is kept constant and $V_{0}$ is changed, the shape of the photoelectric 
curves does not change, but the values of $V_{L}$ corresponding to a given $V_{R}$ increase 
with increasing $V_{0}$. Similarly, if $V_{0}$ is kept constant and the light intensity is changed,
the shape of the photoelectric curves does not change, but the values of $V_{L}$ increase with
increasing light intensity. This behavior is observed for a wide range of high voltages
($V_{0} \sim 700 \div 1100 \units{V}$) and LED intensities (here the range depends on the 
color of the LEDs). However, if the voltage across the load resistor becomes too large 
($V_{L} \gtrsim 1\units{V}$), saturation effects might occur due to the large number of
electrons moving across the last dynodes because the capacitors in the last stages of the 
voltage divider could not be able to keep the voltage differences stable.

Another feature of the photoelectric curves shown in Fig.~\ref{fig:photocurves} 
is that the transition between the regime in which photoelectrons emitted from K 
are collected by Dy1 and the regime in which photoelectrons are repelled  
is not sharp, i.e. the slope of the photoelectric curve changes smoothly with $V_{R}$,
thus making the determination of $V_{S}$ not straightforward. 
This behavior is due to the spread in the photoelectron kinetic energies
when they are emitted from the photocathode.  
It is also worth to point out here that, since photons emitted by LEDs are not monochromatic 
(as shown in Tab.~\ref{tab:LED} the widths of the frequency spectra are $\sim 2 \div 3\%$ of 
the corresponding peak values), the photoelectric curves cannot be described in terms of a 
single value of the stopping potential, but it would be more appropriate to take into account
the dependence of the stopping potential on the frequency. Hereafter we will neglect this
depencence and we will assume that each photoelectric curve can be described in terms of the
stopping potential $V_{S}$ corresponding to the peak frequency of the LED.
 
The determination of either an analytical or a numerical model of the photoelectric curves would 
be rather complex and perhaps would go beyond the scope of an introductory laboratory course for 
undergraduate students. Therefore, to analyze the data collected by the students carrying out 
the experiment, we developed a phenomenological approach. After the analysis of many photoelectric 
curves obtained in different conditions (different LED intensities and different PMT high voltages), 
we noticed that the asymptotic behavior of all photoelectric curves can be adequately described by 
the following functions:

\begin{equation}
 V_{L} = \left\{ 
 \begin{array}{ll}
  a_{1} + b_{1} e^{-\frac{V_{R}}{c_{1}}} & \textrm{for}~V_{R} < V_{S} \\
  a_{2} + b_{2} V_{R} & \textrm{for}~V_{R} > V_{S} \\
 \end{array}
\right.
\label{eq:fit}
\end{equation}
For each photoelectric curve we select two sets of points, belonging to the regions $V_{R}<V_{S}$ 
and $V_{R}>V_{S}$, and we fit these points with the functions in eq.~\ref{eq:fit}, thus determining the 
parameters $a_{1}$, $b_{1}$, $c_{1}$, $a_{2}$ and $b_{2}$. The fits are performed using the free data 
analysis software ROOT~\cite{root}, provided by CERN. We then define the value of the stopping potential 
$V_{S}$ as the abscissa of the intersection point of the two curves, which can be evaluated
solving the following non-linear equation:

\begin{equation}
 a_{1} + b_{1} e^{-\frac{V_{S}}{c_{1}}}  = a_{2} + b_{2} V_{S}.
\label{eq:vsfit}
\end{equation}
The previous equation, which gives $V_{S}$ as a function of the parameters $a_{1}$, $b_{1}$, $c_{1}$, $a_{2}$ 
and $b_{2}$, cannot be solved analytically, but can be easily solved in a numerical way, for instance
using the bisection method. This procedure is graphically illustrated in the plots of
Fig.~\ref{fig:photocurves}, where we superimposed to each photoelectric curve the functions obtained
from the two fits, also showing the position of the intersection between the two curves.
As we anticipated in Sec.~\ref{sec:intro}, it is worth to point out here that the detection of  
a significant steady background current instead of a slowly vanishing current helps to better
define the stopping potential.

To evaluate the error on $V_{S}$ we use the standard error propagation
formula, starting from the errors on  $a_{1}$, $b_{1}$, $c_{1}$, $a_{2}$ and $b_{2}$, which are
computed by the ROOT software when performing the fits. However, since $V_{S}$ is an implicit function 
of the parameters, a numerical approach is also needed to evaluate its partial derivatives with respect to
the various parameters. For instance, to evaluate the partial derivative $\partial V_{S} / \partial a_{1}$,
we start from the set of fitted parameters and we change $a_{1}$ into $a_{1}'=a_{1}+\delta a_{1}$
\footnote{According to the definition of derivative, the condition $\delta a_{1} \rightarrow 0$ must be 
fulfilled, and therefore one must choose $\delta a_{1}$ such that $| \delta a_{1} | \ll |a_{1}| $.}; 
then we solve eq.~\ref{eq:vsfit} with the value of $a_{1}'$, obtaining a new solution $V_{S}'$ and
finally we evaluate the partial derivative as 
$\partial V_{S} / \partial a_{1} \approx \delta V_{S} / \delta a_{1}$,
where $\delta V_{S}=V_{S}'-V_{S}$. In the same way we calculate the partial derivatives of $V_{S}$
with respect to the other parameters. 

\begin{figure}[t]
\includegraphics[width=0.95\columnwidth]{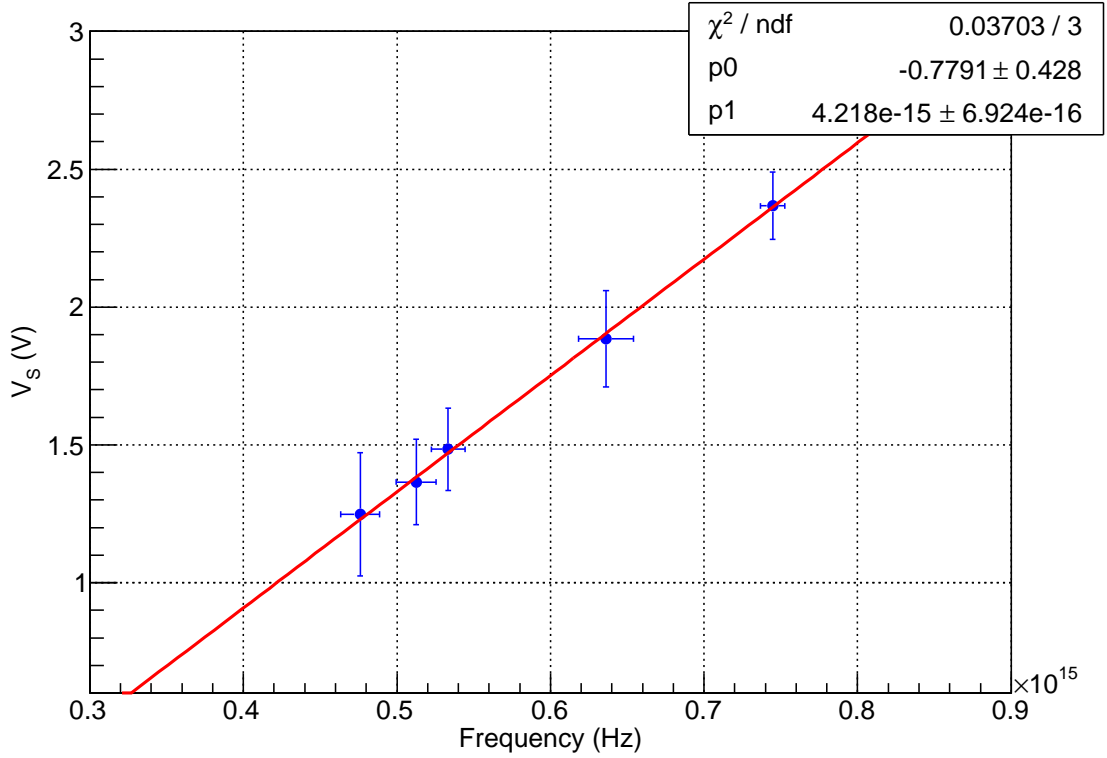}
\caption{Stopping potentials for the various LEDs as a function of frequency.
The horizontal error bars represent the width of the spectra shown in Tab.~\ref{tab:LED},
while the vertical error bars represent the uncertainties on the stopping potentials,
evaluated as discussed in sec.~\ref{sec:analysis}. The points are well fitted with
a straight line. The fit results are summarized in the top panel (the parameters $p_{0}$ 
and $p_{1}$ are respectively the intercept the slope), where the $\chi^{2}$ of the fit 
is also shown.
}
\label{fig:stoppingpotentials}
\end{figure}

The procedure used to evaluate $h/e$ from the measured stopping potentials is illustrated 
in Fig.~\ref{fig:stoppingpotentials}, where the stopping potentials obtained from the analysis of the 
photoelectric curves shown in Fig.~\ref{fig:photocurves} are plotted against the 
frequency of the incident light. The error bars associated to the LED frequencies are 
the widths of their emission spectra, which are taken from 
Tab.~\ref{tab:LED}, while those associated to the stopping potentials are calculated following the approach 
described above. A linear fit of the experimental points is then performed. In the example
shown in Fig.~\ref{fig:stoppingpotentials}, the fit procedure yields a $\chi^{2}/d.o.f.=0.037/3$,
which suggests that the error bars associated to the stopping potentials are overestimated,
a feature which might be a consequence of the phenomenological model that we adopted to describe the
photoelectric curves. According to 
eq.~\ref{eq:vs}, the slope of the line corresponds to $h/e$, while its intercept corresponds to $W/e$.
Assuming for the electron charge the current value $e=1.60 \times 10^{-19}\units{C}$, the fit of the data 
shown in Fig.~\ref{fig:stoppingpotentials} yields for the Planck's constant a value 
$h = (6.75 \pm 1.11) \times 10^{-34} \units{J \cdot s}$ and for the work function of the photocathode 
a value $W = (0.78 \pm 0.43) \units{eV}$.

\section{Discussion and conclusions}
\label{sec:discussion}

The measurement of $h/e$ proposed in the present paper yields an uncertainty of about
$20\%$ on the value of $h/e$ and an uncertainty larger than $50\%$ on $W/e$. 
The main sources of error are the spreads on the LED frequencies and the uncertainties
on the values of the stopping potentials. To mitigate the effects of the frequency spreads, 
one could use monochromatic light sources coupling the LEDs to appropriate filters, 
or even using laser sources. 
%Monochromatic light sources could also allow to obtain 
%photoelectric curves exhibiting sharper slope changes at the position of the stopping potential, making 
%its evaluation easier and perhaps more accurate. 
The uncertainties on the stopping potentials could 
also be reduced with a more appropriate modeling of the photoelectric curves, which goes beyond the
scope of an introductory laboratory course. 

Despite the poor precision attained, we strongly believe that this measurement of $h/e$ is extremely useful
from the educational point of view, because not only it allows to understand the main features 
of the photoelectric effect, but it also stimulates further considerations about the physics involved
in the measurement and on the technique adopted. 
% For instance, when discussing the photoelectric curves in Sec.~\ref{sec:analysis}, 
% we have explained the fact that $V_{L}$ never drops to zero with increasing $V_{R}$ in terms of 
% photons crossing the photocathode and being absorbed by the first dynode. This interpretation will
% allow the student to better understand the concept of photoelectric absorption length, which is
% usually introduced when explaining the construction of photocathodes (see i.e. ref.~\cite{knoll}). 


\begin{thebibliography}{99}

\bibitem{planck} M.~Planck, ``On the distribution law of energy in the normal spectrum'', 
Annalen der Physik {\bf IV}, 7, 553-563 (1901).

\bibitem{einstein} A.~Einstein, ``On a Heuristic Point of View about the Creation and Conversion of Light'',
Annalen der Physik {\bf 17}, 6, 132–148 (1905).

\bibitem{george} S.~George et al., ``Planck's Constant from Wien's Displacement Law'',  
AJP {\bf 40}, 621 (1972).

\bibitem{manikopoulos}  C.~N.~Manikopoulos and J.~F.~Aquirre, ``Determination of the blackbody radiation constant hc/k 
in the modern physics Laboratory'', AJP {\bf 45}, 576 (1977).

\bibitem{crandall} R.~E.~Crandall and J.~F.~Delord, ``Minimal apparatus for determination of Planck's constant'',  
AJP {\bf 51}, 90 (1983).

\bibitem{dryzek} J.~Dryzek and K.~Ruebenbauer, ``Planck's constant determination from black‐body radiation'', 
AJP {\bf 60}, 251 (1992).

\bibitem{brizuela} G.~Brizuela and A.~Juan, ``Planck’s constant determination using a light bulb'', 
AJP {\bf 64}, 819 (1996).

\bibitem{nieves} L.~Nieves et al., ``Measuring the Planck constant with LED's'', 
The Physics Teacher {\bf 35}, 108 (1997).

\bibitem{oleary} A.~J.~O'Leary, ``Two Elementary Experiments to Demonstrate the Photoelectric Law and Measure 
the Planck Constant'', AJP {\bf 14}, 245 (1946).

\bibitem{hall} H.~Hall and R.~P.~Tuttle, ``Photoelectric Effect and Planck's Constant in the Introductory Laboratory'', 
AJP {\bf 39}, 50 (1971).

\bibitem{bobst} R.~L.~Bobst and E.~A.~Karlow, ``A direct potential measurement in the photoelectric effect experiment'', 
AJP {\bf 53}, 911 (1985).

\bibitem{barnett} J.~Dean~Barnett and H.~T.~Stokes, ``Improved student laboratory on the measurement of Planck's constant using the photoelectric effect'', AJP {\bf 56}, 86 (1988).

\bibitem{garver} W.~P.~Garver, ``The Photoelectric Effect Using LEDs as Light Sources'', 
The Physics Teacher {\bf 44}, 272 (2006).

\bibitem{millikan} R.~A.~Millikan, ``A direct photoelectric determination of Planck's h'', 
Phys.\ Rev.\ {\bf 7}, 355-388 (1916). 

\bibitem{millikan2} R.~A.~Millikan, ``The isolation of an ion, a precision measurement of its charge,
and the correction of Stokes's law'', Phys.\ Rev.\ (Series I) {\bf 32}, 349-397 (1911).   

\bibitem{millikan3} R.~A.~Millikan,``On the Elementary Electrical Charge and The Avogadro Constant'',
Phys.\ Rev.\ {\bf 2}, 109-143 (1913).

\bibitem{philips} Philips Photomultipliers: Data Handbook, PC04 (1990). 

\bibitem{ocean} \url{http://www.oceanoptics.com/Products/benchoptions_hr.asp}

\bibitem{caen} \url{http://www.caen.it/csite/CaenProd.jsp?parent=21&idmod=240}

\bibitem{root} \url{https://root.cern.ch/} see also 
R.~Brun and F.~Rademakers, ``ROOT - An Object Oriented Data Analysis Framework'',
Proceedings AIHENP'96 Workshop, Lausanne, Sep. 1996, 
Nucl.\ Inst.\ \& Meth.\ in Phys.\ Res.\ {\bf A 389}, 81-86 (1997). 

\bibitem{knoll} G.~F.~Knoll, ``Radiation Detection and Measurement'',
Wiley (1999).

\end{thebibliography}
\end{document}